\documentclass[10pt,twocolumn]{article}
  \usepackage{amssymb}  
\usepackage{ol2}    
 \usepackage{amssymb} 
 \usepackage{geometry}
\usepackage{graphicx}
\usepackage{epstopdf}

\def\AOM {acousto-optic modulator}
\def\BEC {Bose-Einstein condensate}

\def\CPT {coherent population trapping}
\def\EIA {electromagnetically induced absorption}
\def\ECDL {extended cavity diode laser}

\def\MOT {magneto-optical trap}

\def\ppKTP {periodically poled potassium titanyl phosphate}  

\def\PMT {photo-multiplier tube}

\def\SAS{saturated absorption spectroscopy}

\def\VCO {voltage controlled oscillator}
\def\wrt {with respect to}

\def\ARC {Australian Research Council} 

\newcommand{\degrees}{$^{\circ}$}
\newcommand{\degC}{$^{\circ}$C}
\newcommand{\lwg}{$\Gamma^{(556)}$}
\newcommand{\lwv}{$\Gamma^{(399)}$}

\newcommand{\si}{$\sim$}

\newcommand{\uK}{$\mu$K}

\newcommand{\ee}[1]{\ensuremath{\times 10^{ #1}}}

\newcommand{\fastT}{$^{1}S_{0}-\,^{1}P_{1}$}    
\newcommand{\coolingT}{$^{1}S_{0}-\,^{3}P_{1}$}  

\newcommand{\Yb}{$^{171}$Yb}	
\newcommand{\Ybtwo}{$^{172}$Yb}
\newcommand{\Ybthree}{$^{173}$Yb}
\newcommand{\Ybfour}{$^{174}$Yb}

\begin{document}

  \twocolumn[ 

\title{Inverted crossover resonance aiding laser cooling of  $^{171}$Yb} 

\author{Authors:   J.~J.~McFerran $^{*}$}
\address{$^1$School of Physics, University of Western Australia, 6009 Crawley, Australia}
\address{$^*$Corresponding author: john.mcferran@uwa.edu.au}

\begin{abstract}

We observe an inverted crossover resonance in $\pi$-driven four-level systems, where  $F'-F=0,+1$. 
The signal is observed through saturated absorption spectroscopy  of  the $(6s^{2})$~$^{1}S_{0}$ $-$  $(6s6p)$\,$^{3}P_{1}$  transition in $^{171}$Yb, where the nuclear spin $I=1/2$. The enhanced  absorption signal is used to generate a dispersive curve for 556\,nm laser frequency stabilisation and the stabilised light  cools $^{171}$Yb atoms in a two-stage magneto-optical trap,  achieving temperatures of 20\,$\mu$K.    The  Doppler-free spectroscopy scheme is further used to measure isotopic frequency shifts  and hyperfine separations for the intercombination line in ytterbium.

\end{abstract}


\maketitle 

\vspace{0.5cm}

    ] 

\section{Introduction}

Neutral ytterbium  finds a niche in many ultra-cold atom explorations; examples include degenerate Fermi gases~\cite{Fuk2007a,Han2011,Dor2013},  artificial gauge potentials~\cite{Dal2011,Tai2012, Sca2014}, quantum many body simulations~\cite{Bra2013b,Hof2015} and atomic clocks~\cite{Por2004a,Hin2013,Bel2014,Tak2015a,Nem2016}, along with the potential application in quantum computation~\cite{Dal2011a}.  In the majority of instances the intercombination line (\coolingT) is used for  laser cooling, typically in a magneto-optical trap. 
With a Doppler cooling limit  of 4.5\,\uK\ the low temperatures are conducive to a range of applications; for example, loading optical lattice traps and from there producing all optical \BEC s~\cite{Tak2003b} or degenerate Fermi gases~\cite{Fuk2007a,Dor2013}.    With a natural line-width of 184\,kHz the full potential of the laser cooling is only attainable if the frequency noise of the 556\,nm light is sufficiently low.  One solution is to use an optical cavity with a reasonably high finesse and low frequency drift to stabilise the 556\,nm light frequency~\cite{Kuw1999,Mar2003b}.   
 An alternative is to lock the green light  to an element of  a well stabilised frequency comb~\cite{Yas2010}.
 A further  means is to lock to the center of a Doppler profile produced with  a thermal atomic beam~\cite{Tas2010,Hoy2005}. This helps maintain long term stability of the green light; however, keeping the frequency excursions close to 180\,kHz is challenging~\cite{comment1}.  
Here we demonstrate a means of conferring short-to-long term frequency stability on 556\,nm light without the need for an optical cavity or frequency comb lock. 
We use frequency-modulated spectroscopy where  the 556\,nm beam intersects a Yb atomic beam at normal incidence, and reflect the beam with a lens-mirror combination to create a pump-probe scheme for fluorescence spectroscopy.  
Unexpectedly, we observe  enhanced absorption (rather than saturation) behaviour in \textit{saturated} absorption spectroscopy (SAS) for both of the hyperfine lines in the \coolingT\ transition of \Yb.   The effect occurs when a small DC magnetic field, of the order of a few hundredths of a millitesla, is applied across the interaction zone. The field strength is below that required to resolve the Zeeman sub-levels; though, is still apparent when the Zeeman sub-levels are separated.   We have identified the enhanced absorption as an inverted crossover resonance between two principal transitions~\cite{Han1971,Pap1980,Nak1997}, where the principal transitions arise due to the Zeeman splitting of magnetic sub-states.
The signal we observe is similar to predictions made by models of  \EIA\ (EIA) such as that by Taichenachev \textit{et al.}~\cite{Tai1999} and Goren \textit{et al.}~\cite{Gor2004}.
However, the pump-probe scheme with linearly polarised light  that is used in the experiment here does not permit a \textit{transfer of coherence} as required for EIA. 

   Doppler-free spectroscopy in Yb for frequency stabilisation of 556\,nm light has been reported previously, but without reference to detail~\cite{Dor2013}.  We also note that there is recent discussion about \CPT\ (CPT) with the intercombination line in \Ybfour~\cite{Sin2015a,Vaf2016}.   A characteristic that we share with these CPT investigations is that  there is saturation of the transition by both pump and probe fields. 
   Our observation of an inverted crossover resonance may be the first for a group II atom. 

The instability of the 556\,nm light frequency, locked to the crossover resonance, is assessed by use of an optical frequency comb whose mode spacing  is referenced to a hydrogen maser.   
Further validation of the the green light's frequency  instability is performed by using the 556\,nm light in a two-stage \MOT\ of \Yb\ ($F'=3/2$), producing temperatures below  20\,\uK\ with a standard deviation  of 2\,\uK\ across nearly sixty sequential measurements ($F'$ represents the total angular momentum of the exited state). 

Our setup also enables a  measure of frequency differences between isotopic lines and hyperfine separations with statistical uncertainties of 60\,kHz or less for most of the frequency intervals. 
  These are determined via absolute optical frequency measurements~\cite{Nen2016}. We provide a summary of these measurements and compare with previous reports.  
 Relatively good agreement is found,   though a number of frequency separations differ by more than two standard deviations.  High-resolution spectroscopy on the Yb intercombination line has previously been performed via $V$-type double-resonance spectroscopy, but without quantitative frequency measurements~\cite{Lof2001a}.   

   This paper is organised as follows.  Section 1 describes the experimental details for the \SAS\ (SAS) and frequency measurements, Sect. 2  presents the spectra generated from the \SAS\ and gives information pertaining to the enhanced absorption, Sect. 3 describes the two-stage laser cooling scheme of \Yb\ and presents results related to the stabilisation of the 556\,nm light, and Sect. 4  summarises the isotopic frequency shift measurements carried out with the saturated absorption scheme.

\section{Saturated absorption spectroscopy}

\begin{figure}[h]			
 \begin{center}
\resizebox{0.5\textwidth}{!}  
{		
  \includegraphics[width=7.5cm,keepaspectratio=true]{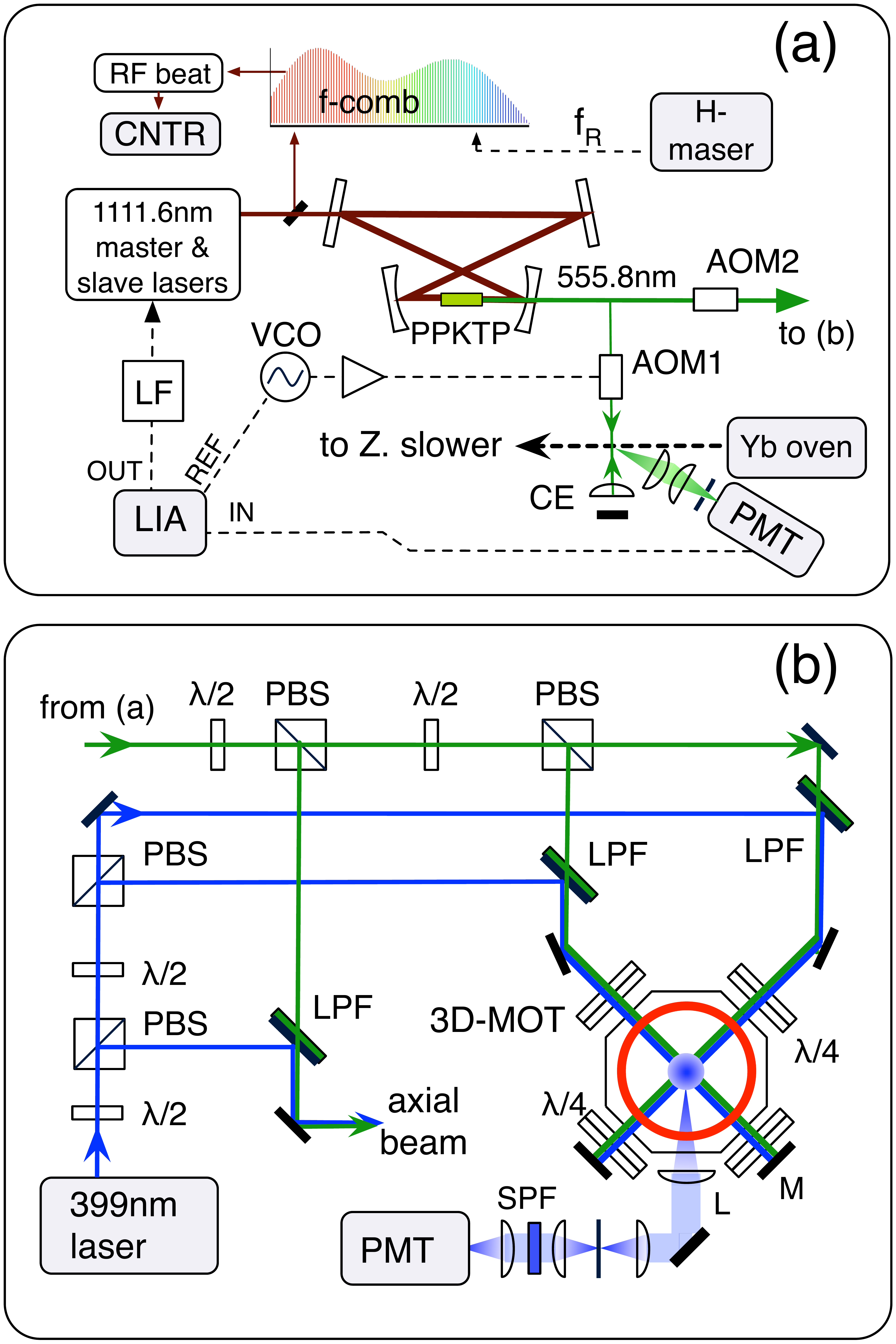}}   
\caption[]{\footnotesize     
(a) The principal parts for the saturated absorption spectroscopy of the \coolingT\ transition in an atomic beam of Yb.  (b) The experimental arrangement used for two-stage cooling of Yb in a $\sigma^{+}-\sigma^{-}$ \MOT\ after laser stabilisation to the saturated absorption signal of $^{171}$Yb $(F'=3/2)$.
 AOM, \AOM; CNTR, frequency counter; CE, cat's eye reflector; ECDL, \ECDL; f-comb, frequency comb (in the near-IR); $f_{R}$, repetition frequency;  H-m, hydrogen maser;  LF, loop filter (proportional and integral gain); LIA, lock-in amplifier; LPF, long-wavelength pass filter; MOT, \MOT; PBS, polarizing beam splitter; PMT, \PMT; PPKTP, \ppKTP;  SPF, short-wavelength pass filter;  VCO, \VCO; Z, Zeeman.  
  }
   \label{SASandMOT}  %
\end{center}
\end{figure}
%

  The saturated absorption scheme comprises a retro-reflected 556\,nm beam at right angles to the atomic beam and the 556\,nm fluorescence is detected by a \PMT\ (Hamamatsu H10492-003).   We  take advantage of the Yb atomic beam leading to a \MOT\ (MOT) used for clock line spectroscopy \cite{Nen2016}.   Frequency modulation at 33\,kHz is applied to the light via an \AOM\ (AOM) allowing for third harmonic detection with a lock-in amplifier.   Helmholtz coils in close vicinity of the saturated absorption zone are used to set the magnetic field in the vertical direction.   For isotopic frequency shift measurements, the magnetic field is set to cancel the vertical component of the Earth's $\textbf{\emph{B}}$-field (0.059\,mT upward), and the polarisation of the 556\,nm light is set  vertically (i.e. parallel to that of $\textbf{\emph{B}}$ when it is non-zero).   The zero $B$-field condition has been determined by several means; one of which is to rotate the light polarisation by 90\degrees\ and observe the Zeeman splitting, which is stronger for the $\sigma$ transitions.  
  The intensity profile of the 556\,nm beam is elliptical with the major axis aligned with the atomic beam.  
  
The temperature of the oven (Createc, GmbH) is maintained at 430\degC\ producing an effusion of Yb atoms through a trio of stainless steel tubes with a diameter of 1\,mm and length 20\,mm.  
  The green  fluorescence passes through a spatial filtering scheme to minimise stray light.   We estimate the flow rate of Yb atoms through the 556\,nm beams to be \si4(3)\ee{13}\,s$^{-1}$  based on molecular flow rates through a cylindrical tube~\cite{Ram1956} (an estimate based on fluorescence detection is an order of magnitude less).  Based on a mean velocity of 320\,m\,s$^{-1}$ and an intensity well above saturation, the number of absorption/emission cycles is about 15 per atom as the atoms traverse  the pump and probe beams.   
   The 556\,nm light is produced via resonant frequency doubling of 1112\,nm light, which is generated from a commercial fiber laser that injection-locks a diode laser~\cite{Kos2015}.  A bow-tie configuration doubling cavity uses an 18\,mm length \ppKTP\ crystal for the nonlinear conversion and 75\,mm radius of curvature focusing mirrors.  The line-width of the 556\,nm radiation has been found to be no greater than 410\,kHz~\cite{Kos2015}. Based on the manufacturer specifications for the fiber laser, the line-width of the green light should be below 240\,kHz $-$ slightly larger than the natural line-width of the intercombination line. 

For generating the spectra the 556\,nm light frequency is slowly swept via a piezo transducer in the fiber laser. We calibrate the frequency span by use of a frequency comb.  There is 0.8\, mW of 1112\,nm light combined with the nearest comb element on an avalanche photodiode  to produce an RF beat signal with 29\,dB signal-to-noise ratio (SNR) with 500\,kHz resolution bandwidth.  The beat signal is filtered with a 10\,MHz wide ($-3$\,dB) band pass filter centered at 30\,MHz. The frequency comb, referenced to a hydrogen maser,  is used to make frequency measurements for all of the Yb isotopic lines, except for $^{168}$Yb with 0.13\,\% relative abundance  and  $^{173}$Yb\,$(F'=3/2)$ (Sect.~\ref{Isotope}).   



   \begin{figure}[h]
 \begin{center}
\resizebox{0.5\textwidth}{!}  
{		
  \includegraphics[width=8.8cm,keepaspectratio=true]{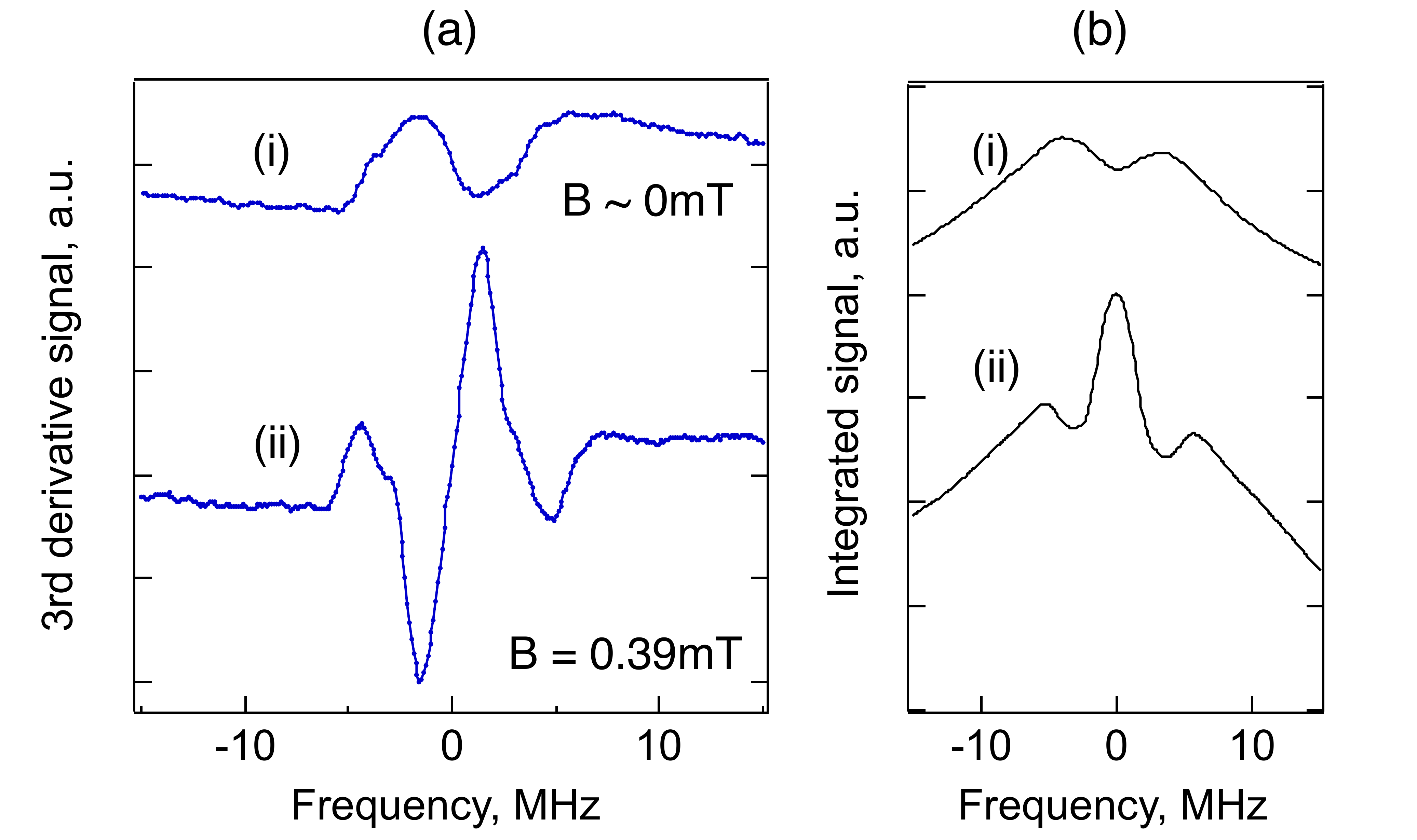}}   %
\caption[]{\footnotesize     
(a) Third harmonic dispersive curves for saturated absorption spectroscopy of the $^{1}S_{0}-\, ^{3}P_{1}$ line of $^{171}$Yb ($F'=1/2$) for two values of magnetic field:  (i) near zero field  and (ii) $B=0.39$\,mT.   A vertical offset is applied for clarity.  The result in trace (ii) occurs for both directions of the applied $B$-field.  (b) The integrated dispersive profiles of (a).  The averaging time per data point was 0.2\,s. 
 }
   \label{EnhancedAbs}  %
\end{center}
\end{figure}

A comparison of the dispersive spectra for $^{171}$Yb ($F'=1/2$) at two different magnetic field strengths is shown in Fig.~\ref{EnhancedAbs}.  Trace (a) with  zero $B$-field shows the usual saturated absorption feature (or Lamb dip), while at \si0.39\,mT  there is a strong inversion of the dip showing  enhanced absorption.   The enhancement is seen for both magnetic field directions and for both  $^{171}$Yb  hyperfine lines, however the effect is more pronounced with $^{171}$Yb ($F'=1/2$). 
The strength of the enhanced absorption is reduced by about 30\,\%  in the case of $^{171}$Yb ($F'=3/2$). 
 The difference between the two traces shows that there is a substantial change in behaviour with the simple addition of a small DC magnetic field. 
 
\begin{figure}[h]
 \begin{center}
\resizebox{0.5\textwidth}{!}  
{		
  \includegraphics[width=7.2cm,keepaspectratio=true]{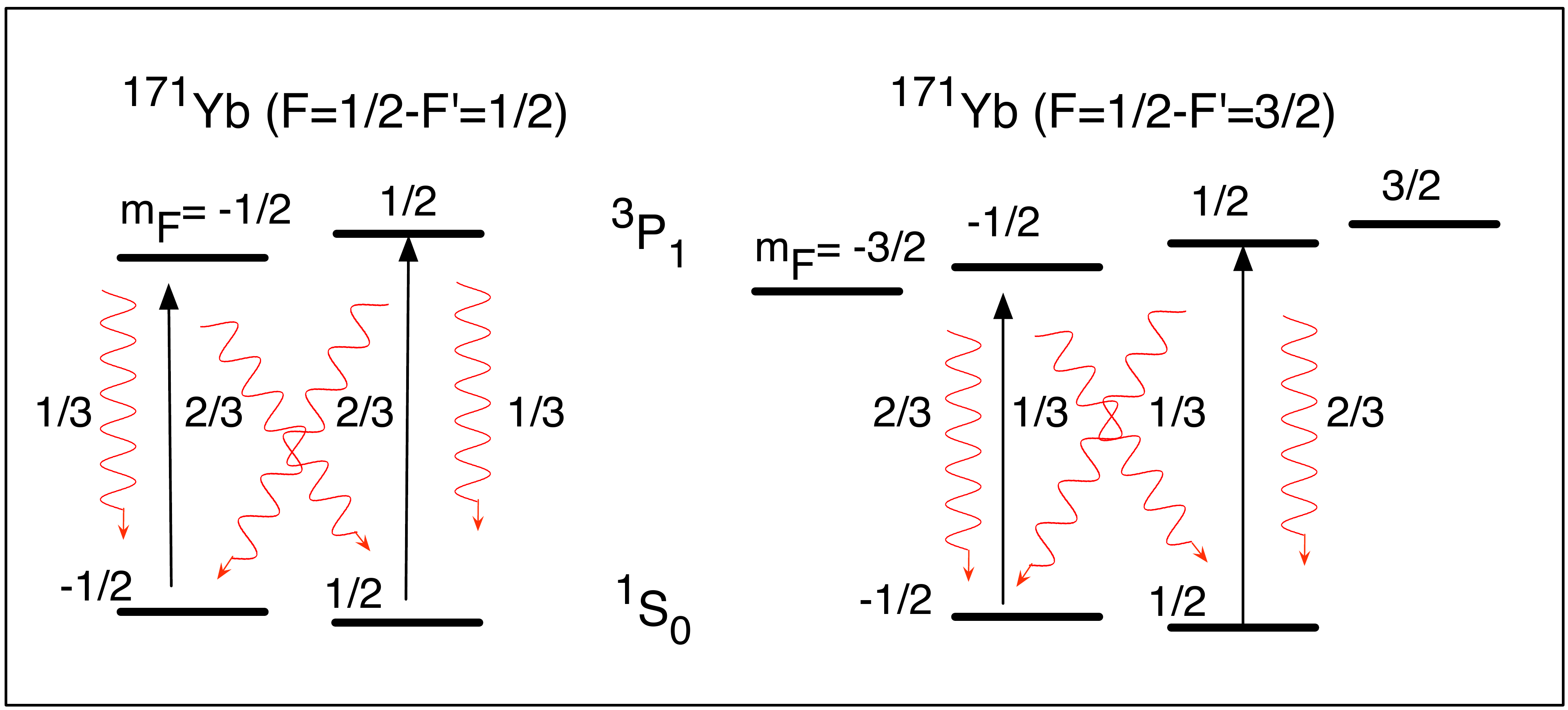}}   %
\caption[]{\footnotesize     
Branching ratios from the upper state Zeeman sub-levels in \Yb\ for the $F=\frac{1}{2}-F'=\frac{1}{2}$ and $F=\frac{1}{2}-F'=\frac{3}{2}$ transitions.   For the excitation phase the 556\,nm light polarisation is parallel to the magnetic field, implying $\Delta m_{F}=0$.  The separation of the upper and lower states is not drawn to scale.  The sub-level separation rates are: 14\,MHz$\cdot$mT$^{-1}$ for $F=\frac{1}{2}-F'=\frac{1}{2}$,   7\,MHz$\cdot$mT$^{-1}$ for $F=\frac{1}{2}-F'=\frac{3}{2}$   and 3.76\,kHz$\cdot$mT$^{-1}$ for the ground state.
 }
   \label{mFlevels}  %
\end{center}
\end{figure}

\begin{figure}[h]
 \begin{center}
{		
  \includegraphics[width=8.4cm,keepaspectratio=true]{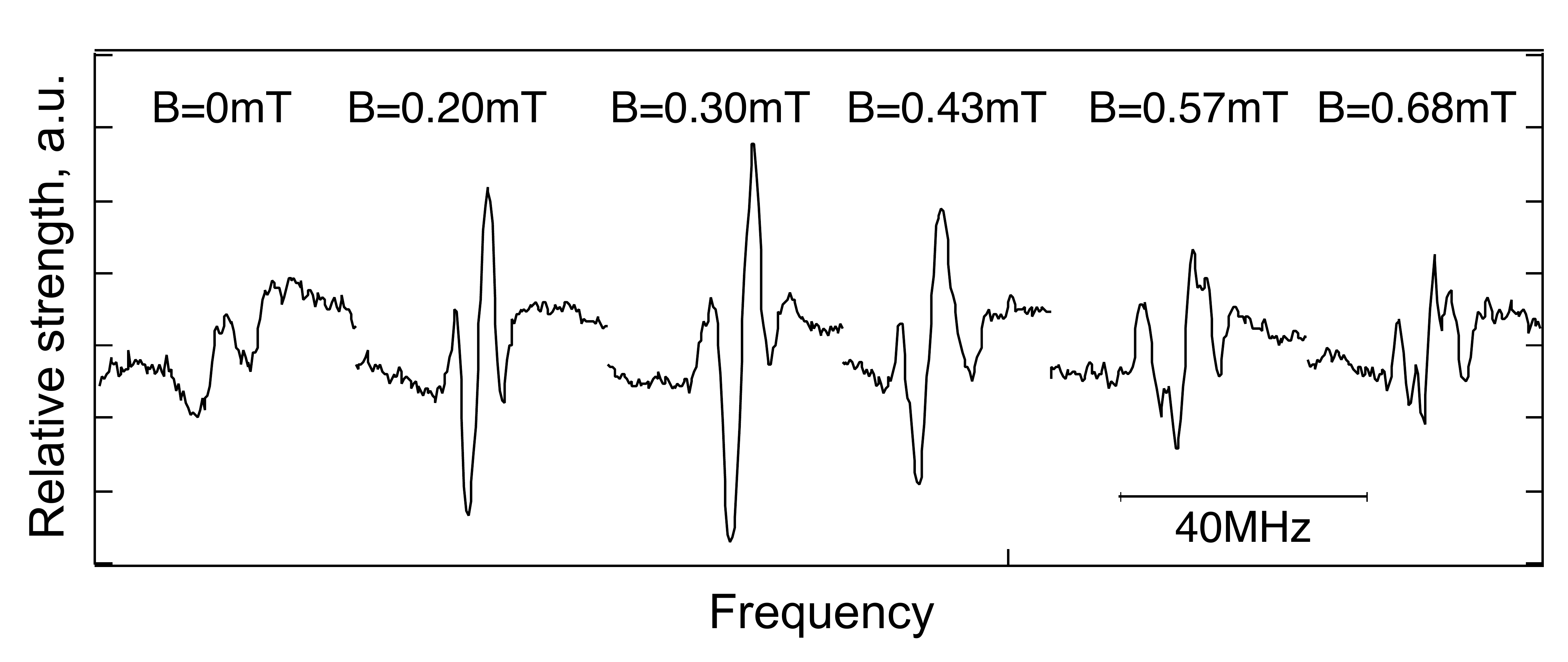}}   %
\caption[]{\footnotesize     
The dispersive curves produced through saturated absorption for \Yb\ ($F'=1/2$) with different levels of bias magnetic field.   The Zeeman sub-levels of the upper state begin to appear at the higher $B$-field strengths. 
 }
   \label{SpectravsB}  %
\end{center}
\end{figure}

  The branching ratios for the de-excitation are readily calculated for the $^{171}$Yb ($F'=1/2$)  and $^{171}$Yb ($F'=3/2$) systems~\cite{McC1996}.  We illustrate these in Fig.~\ref{mFlevels}.    For $^{171}$Yb ($F'=1/2$)  the probability of the downward $\sigma$ transitions is 2/3 while for $^{171}$Yb ($F'=3/2$)  it is 1/3.  This correlates with the effect being more pronounced for $F'=1/2$.   The factors relevant here are the natural line-width of the transition (\lwg\,$=184$\,kHz), the Zeeman shift of the $^{3}P_{1}$  $m_{F}$ levels ($\pm14$\,MHz$\cdot$mT$^{-1}$) and the Zeeman shift of the ground state $m_{F}$ levels ($\pm3.76$\,kHz$\cdot$mT$^{-1}$).  
  The latter is evaluated with $\Delta\omega=g_{F}\mu_{B} m_{F}B/\hbar$, where  $ g_{F}$ is the Land\'{e} $g$-factor and $\mu_{B}$ is the Bohr magneton. For the case of hyperfine states in an external magnetic field with $J=0$ and $F=I$ the  Land\'{e} $g$-factor simplifies to become
 
 \begin{equation}
\label{gfactor}
g_{F} =g_{I} = -\frac{1}{I}\left(\frac{\mu_{N}}{\mu_{B}}\right)\left(\frac{\mu_{I}}{\mu_{N}}\right)
\end{equation}
where the first quotient is the ratio of nuclear and Bohr magnetons (hence, equal to the ratio of electron and proton masses) 
 and the second quotient is the nuclei magnetic moment expressed in nuclear magnetons~\cite{Der2011a,Hak2000}. 
  For $\mu_{I}/\mu_{N}$ we take the value of 0.49367(1) from published tables~\cite{Sto2014}. 
The ground state Zeeman splitting is $2.7\times10^{-4}$ times weaker than the upper state splitting.  This degree of splitting may not be relevant here, but is included as a matter of interest. 
The Rabi frequency for the pump is \si1.2\,MHz, while that for the retro-reflected  beam is approximately 5\,\%\ less.  
  We note that the \Yb\ crossover resonance  occurs before the Zeeman sub-levels become resolved, but is still present as the pair of Zeeman lines separate when $B \gtrsim 0.5$\,mT, as seen in Fig.~\ref{SpectravsB}, which was recorded for the $F'=1/2$ line.  
  

The enhanced absorption signal is best described as an inverted crossover resonance, where the principal transitions are those depicted in Fig.~\ref{mFlevels} and are driven with $\pi$-polarised light.  Two subsets of atoms are involved, each with a velocity component of equal magnitude but with opposite directions in regard to zero transverse velocity (as defined by the direction of the 556nm beams). 
When the atom has four levels, an atom moving obliquely  can interact with the two beams. 
 Each set of atoms interacts with both the forward and reverse beams, exciting one or the other transition depending upon its direction of transverse motion.   The combination of two sets of atoms usually creates a stronger absorption signal than that generated by purely zero velocity atoms.  
  For the \Yb\ transitions, in addition to this, the branching to the ground state doublet creates multiple pathways for the atoms through the four states.  With multiple pathways the system is no longer saturated; it permits free cycling of the atoms.  The photon absorption is now only limited by the interaction time, hence the enhancement in the absorption. 




The modelling carried out by Vafafard \textit{et al.}~\cite{Vaf2016} is partly relevant given that they studied the intercombination of ytterbium for a saturated pump-probe scheme; however, their analysis was for $^{174}$Yb, which has a non-degenerate ground state.  We believe the enhanced absorption signal observed in the experiment here is due to a four-level dynamic rather than three, because the signal contrast is different for the two hyperfine lines with different sub-level branching ratios.  

  Enhanced absorption (or an inverted Lamb dip) has been observed previously in SAS; for example, in $^{133}$Cs~\cite{Sch1994}  and $^{40}$Ca~\cite{Ish1994}.  In the former case  the enhanced absorption occurs with a change in $F$ or in crossover resonances.  For example, the saturation behaviour depends strongly on the magnetic field for  the $F=2\rightarrow F'=2,3$ crossover resonance.   The level system in \Yb\ is simpler with only four Zeeman sub-states involved.   We may be reporting the first example of a crossover resonance where $F=F'$. 
  
  For $^{40}$Ca the inverted dip occurs for a particular range of intensity levels and forms a triplet at high intensity. In our case the enhanced absorption is manifest for the 556\,nm  beam intensity range of 18\,$I_{\mathrm{S}}^{(556)}$ to 910\,$I_{\mathrm{S}}^{(556)}$, where  $I_{\mathrm{S}}^{(556)}=1.4$\,W\,m$^{-2}$ is the saturation intensity for the \coolingT\ transition (we have stated the combined intensity of the two beams).  The only changes that occur to the line are an enhanced SNR with some line broadening, though both appear to become constant above about 150\,$I_{\mathrm{S}}^{(556)}$.   Below 18\,$I_{\mathrm{S}}^{(556)}$ the signal is too weak to discern, so it is not clear what happens near saturation.    

 
For the \Yb\ ($F'=3/2$) hyperfine transition  with a nulled magnetic field, a very small perturbation to the line is seen that  we attribute to the \Ybthree\ ($F'=3/2$) line.   It is consistent with previous reports  that the latter is within  3\,MHz of the \Yb\ ($F'=3/2$) line and higher in frequency~\cite{Cla1979,Pan2009}. 

An inverted crossover resonance is again observed for $^{173}$Yb\, $(F=\frac{5}{2}\rightarrow\frac{5}{2})$, however the contrast is significantly weaker than for \Yb. The spin-$\frac{5}{2}$ system has a weaker Zeeman shift dependence (1.2\,MHz$\cdot$mT$^{-1}$). Observations have been made for magnetic fields up to 1.2\,mT without a high contrast resonance occurring.  
The higher spin system generates an abundance of transition pathways, which likely prevents the observation of a clear crossover resonance since the multi-level splitting encompasses a broader atomic velocity range.

A crossover resonance is also observed for \Ybtwo\ (and by inference all the bosonic isotopes) when $\textbf{\emph{E}}$ and $\textbf{\emph{B}}$ are perpendicular, thus driving $\Delta m_{F}=\pm1$ transitions.  In this case the resonance exhibits reduced absorption (i.e., it is a normal Lamb dip). Here, the two atomic velocity groups share only two transition pathways.  


The enhanced absorption signal of $^{171}$Yb\,($F'=3/2$) is used to stabilise the frequency of the 556\,nm light  by locking to the center of the dispersion curve. The correction signal is sent to the piezo transducer in the 1112\,nm master fiber laser with a servo bandwidth of a few Hertz.  The instability of the 556\,nm light frequency was assessed by counting the beat signal frequency generated by the heterodyning of 1112\,nm light with an adjacent element of the frequency comb.  As mentioned, the mode spacing (and offset frequency) of the comb were referenced to a hydrogen maser.    The instability at 1\,s integration time is approximately 70\,kHz (for the carrier frequency corresponding to 556\,nm) and integrates down as $\tau^{-1/2}$, reducing to 6\,kHz after 100\,s of integration time.

\section{Laser Cooling}

 This stabilised 556\,nm light is used in a two-stage laser cooling scheme, where atoms are first trapped with 399\,nm light and a quadrupole magnetic field;  then for 50\,ms the violet light is switched off and the atoms are further cooled with the 556\,nm light.   
  The dual-MOT scheme is depicted in Fig.~\ref{SASandMOT}(b).  The 399\,nm and 556\,nm laser beams are combined with long-wavelength pass filters for each of the MOT beams, and achromatic quarter wave plates are used to set the handedness of the circular polarisation (simultaneously for both wavelengths). 
  Since the 399\,nm light is responsible for the initial capturing of the Yb atoms, the 556\,nm beam diameter in the MOT can be set small, so posing less need for high power (through,  at the expense of more stringent alignment conditions).  The  556\,nm beam radius ($e^{-2}$) is 2.0\,mm and approximately 9\,mW in total from the six beams reaches the atoms. 
  
 The difference frequency of  RF signals driving two  \AOM s is used to set  the frequency detuning of the 556\,nm MOT light, where  one AOM shifts the frequency of the light sent to the saturation absorption scheme and the other the MOT light.  The transfer efficiency is found  to be highest for a frequency detuning of $-6.1$\,MHz or about $-33$\,\lwg.

 
 \begin{figure}[h]
 \begin{center}
\resizebox{0.5\textwidth}{!}  
{		
  \includegraphics[width=8.0cm,keepaspectratio=true]{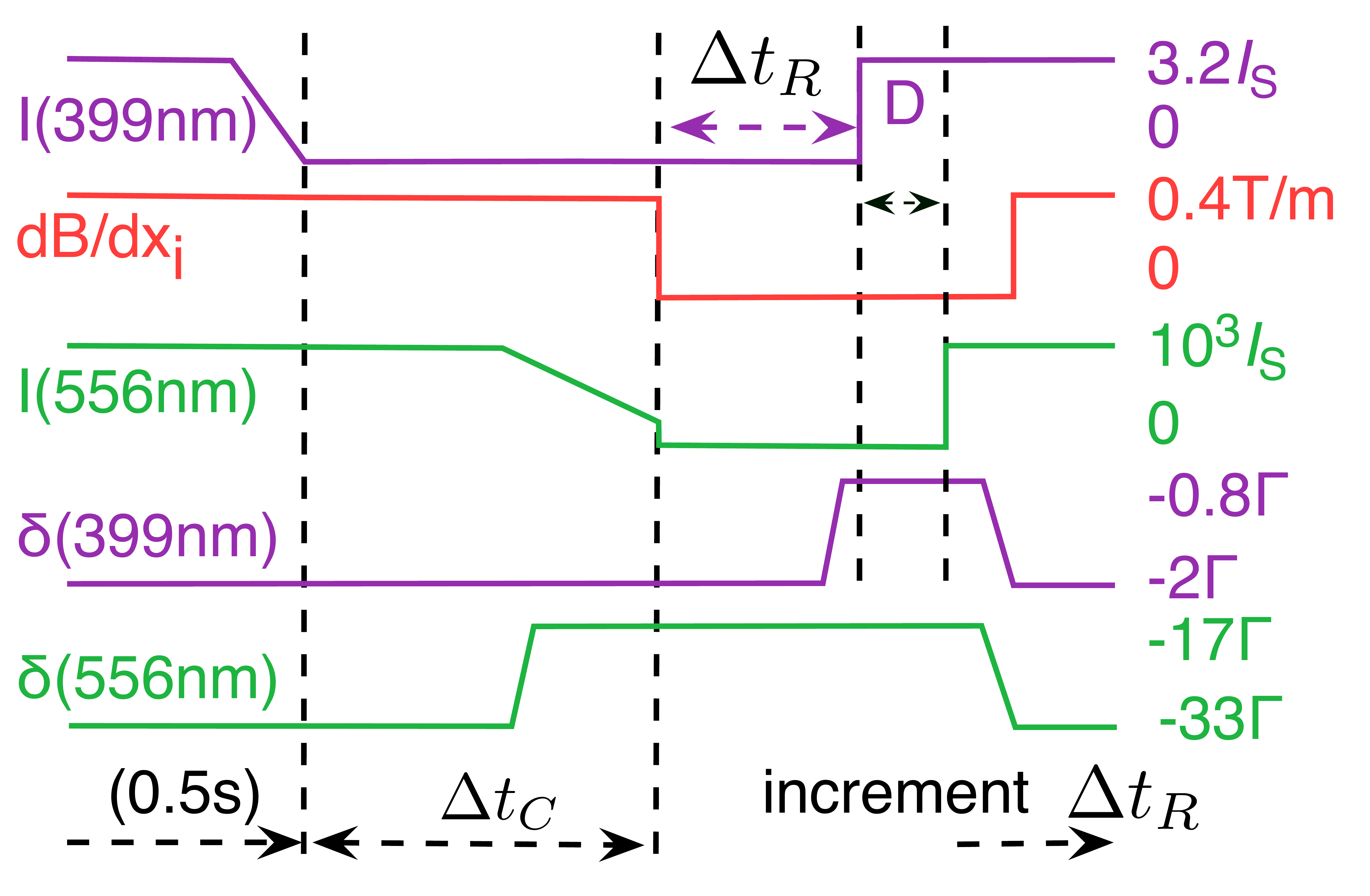}}   
\caption[]{\footnotesize     
 Event sequence for temperature measurements with the two-stage \MOT.  $I$, intensity; $D$, detection period; $\Delta t_{R}$, free expansion duration;  $I_{S}= 595$ W\,m$^{-2}$ and 1.4\,W\,m$^{-2}$ for  the \fastT\ and \coolingT\ transitions, respectively; $\delta$ represents the frequency detuning from line-center.  \lwv\ $= 2\pi \times 28.9\times 10^{6}$\, rad\,s$^{-1}$ 
 and \lwg~$= 2\pi \times 184\times 10^{3}$\, rad\,s$^{-1}$.  The additional cooling phase produced by the 556\,nm light occurs over the period $\Delta t_{C}$,  set to 50\,ms. $\Delta t_{C}$ and $\Delta t_{R}$ are not drawn to scale.   
  }
   \label{SequenceYb}  %
\end{center}
\end{figure}

\begin{figure}[t]
 \begin{center}
\resizebox{0.5\textwidth}{!}  
{		
  \includegraphics[width=8.0cm,keepaspectratio=true]{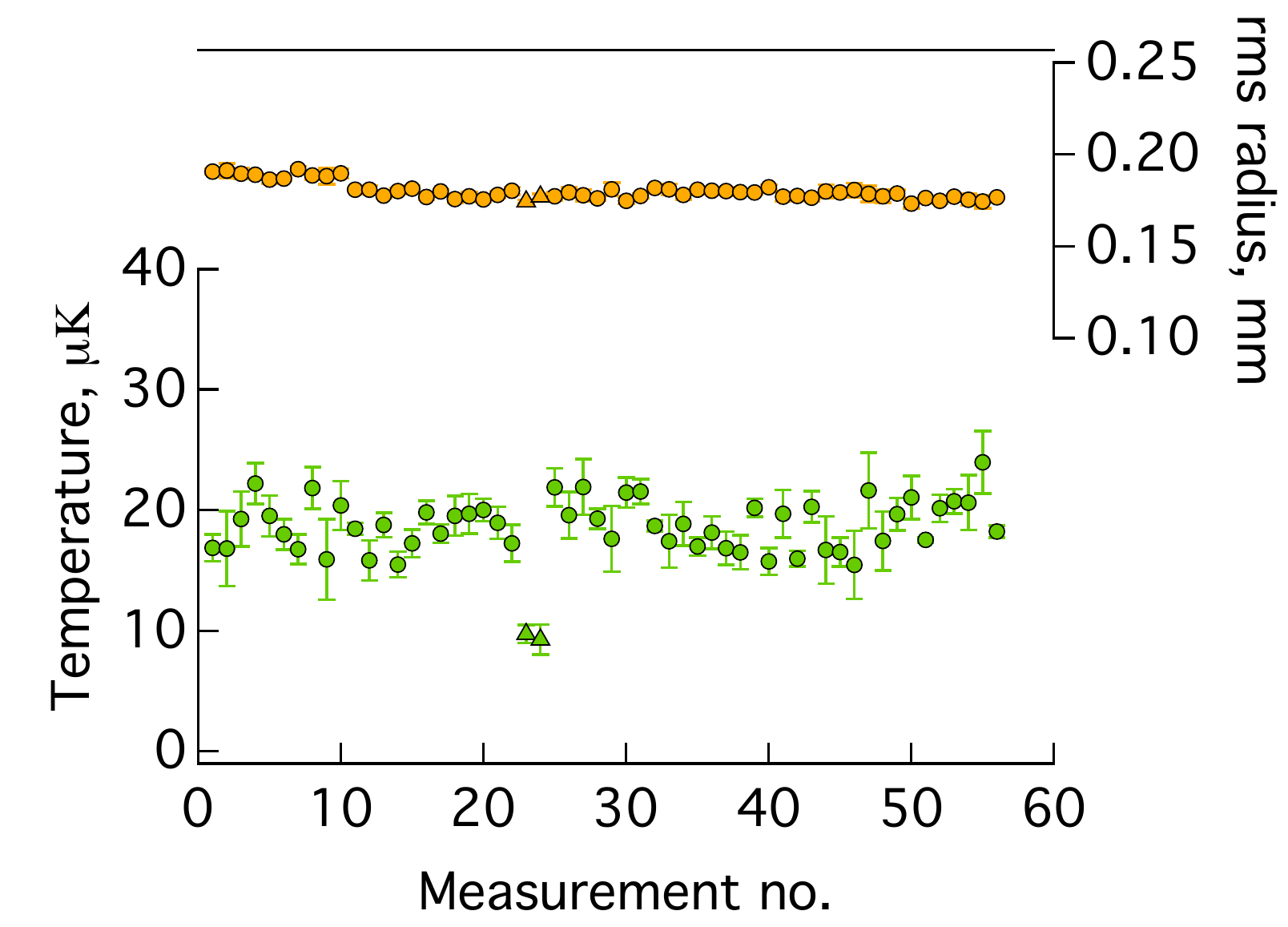}}   
\caption[]{\footnotesize     
Temperature and MOT cloud radius recorded for $^{171}$Yb ($F'=\frac{3}{2}$) using the  inverted crossover resonance to stabilise the 556\,nm light.  The weighted mean temperature for this sequence is $18.2\pm2.0$\,\uK. 
  The two triangular data points show the effect of not releasing the atoms from the magnetic quadrupole field for the free expansion period.   
  }
   \label{Temperature}  %
\end{center}
\end{figure}

Temperature measurements were performed by recording  the free expansion of the atom cloud on a CCD; each successive image is generated through the reloading  and release of the MOT~\cite{Kos2014} (the initial release time was 1\,ms and incremented in 0.75\,ms steps for a series of five frames, the CCD exposure time was 1\,ms).   The optical imaging and filtering scheme produces a magnification of unity.  
 The event sequence for the dual-colour MOT  is summarised in Fig.~\ref{SequenceYb}, and 
  Fig.~\ref{Temperature} shows a succession of temperature and cloud radius measurements.  Here the 399\,nm light was switched off for 50\,ms before the release of the remaining electromagnetic fields. Apart from the free expansion period, the magnetic quadrupole field  was held fixed at 0.42\,T$\cdot$m$^{-1}$ (axial direction).  This value was calculated based on the   design of the MOT coils and current. 
  We found that a reduction in $B$-field gradient for the 556\,nm MOT loading did not aid in the transfer (further discussion below). 
    The frequency detuning, $\delta$, of the 556\,nm light was at a  $-33$\,\lwg\  for the transfer and then swept to   $-17$\,\lwg\  in concurrence with a reduction in intensity to \si200\,$I_{S}^{(556)}$, from a maximum of  $10^{3} I_{S}^{(556)}$.  
     Ramping down the intensity of the 556\,nm light  is found to have influence only when used in combination with a reduction in $\delta$.   
 The uncertainties given by the error bars are generated from the line fitting of  $r_{\mathrm{rms}}^{2}$ versus $t^{2}$ produced by the free expansion of the atom cloud. The weighted mean temperature is 18.2\,\uK\ and the fluctuation in temperature, as represented by the standard deviation of the data set, is 2.0\,\uK.  
 The low level of temperature variation from shot to shot  (or cycle to cycle) is a good indication of the stability of the 556\,nm laser frequency and hence the merit of locking to the inverted crossover resonance. 
 Reducing the green light intensity closer to $I_{S}^{(556)}$  has yet to produce lower atomic cloud temperatures (where the trapping force is still about ten times greater than that due to gravity).   The 18\,\uK\ temperature measured here is comparable to that achieved in other Yb MOTs that use the intercombination line (e.g. Ref.~\cite{Nem2016} reports 15\,\uK). 
 For reference,   the effect of releasing the atoms from the light fields only and not the quadrupole $B$-field is indicated by the two triangular data points in Fig.~\ref{Temperature}.

   We note that the 556\,nm light is shut off via switching of the AOM's RF drive. This leaves a residual 0.05\,$I_{S}^{(556)}$  of light intensity at the MOT.  We confirmed that this did not influence the temperature measurements by use of a physical shutter (Cambridge Technology) to switch the 556\,nm beams on and off~\cite{comment2}.  No change in the temperature recordings was evident. The limitation preventing lower temperatures  is still to be determined  (as previously noted, the limit set by 1D Doppler cooling theory is 4.5\,\uK).  A likely influence is the Zeeman slower beam, which remains on continuously and is only detuned by $-7$\,\lwv\ from the center of the \fastT\ transition.   However, variations of the Zeeman beam intensity by 50\,\%  (from  \si2.5\,$I_{S}^{(399)}$ to  \si5\,$I_{S}^{(399)}$) show no discernible influence on the atomic cloud temperature.   Based on ion pump current readings the chamber pressure was approximately    5\ee{-8}\,Pa  (or 5\ee{-10}\,mbar).  
  Remaining higher frequency noise on the green light is a possible reason given that we rely on the intrinsic line-width of the 1112\,nm master laser and servo bandwidth of the resonant frequency doubling cavity.   The  556\,nm MOT beams are also  prone to very slight direction changes  as the frequency detuning is reduced.   
    We have also studied the temperature of the composite bosonic isotope \Ybtwo\  with the same event sequence. 
     Here there is no enhanced absorption signal in the SAS, so the laser stabilising is with respect to the more usual Lamb-dip signal.   The combination of reducing $\delta$ and the 556\,nm intensity before the free expansion is very pronounced. 
 We observe a temperature reduction from 110\,\uK\ to 45\,\uK\ through this additional cooling phase, which is carried out over 15\,ms.  

     The cloud radius (root-mean squared value) in Fig.~\ref{Temperature} gives a mean value of about  \si0.18\,mm.  
     From the fluorescence received by a \PMT\ (or with the CCD, calibrated) we can infer the number of atoms in the MOT.  The collection efficiency is about 0.5\,\%, and the PMT sensitivity is 0.085\,V\,nW$^{-1}$ for the choice of gain setting. Taking into account the frequency detuning and light intensity we estimate the number of \Ybtwo\ atoms to be 2\ee{6}, corresponding to a (peak) density of 3\ee{10}\,cm$^{-3}$.  
     This number reduces by 40\,\% in the transfer from the violet to green MOT and the number of \Yb\ atoms trapped is \si40\,\% that of \Ybtwo.   The low atom number is attributed to the low atomic flux rate from the oven, due to the conservative nozzle design.   As the need arises this nozzle will be replaced such that the flow rate is enhanced. 
      A higher flux rate is not anticipated to influence the atomic cloud temperature significantly (the present flow rate of remaining thermal atoms through the MOT cloud is roughly 1\ee{9}\,s$^{-1}$). 

\begin{table*}
 \caption{Summary of the measured \coolingT\ isotopic shifts and hyperfine splittings in ytterbium.
Some previously reported values are included for comparison. \label{isoshifts}}
 \begin{tabular}{lllll}
    \vspace{-0.3cm}  \\
  \hline  
     \vspace{-0.3cm}  \\
 &  & Shift from  $^{176}$Yb (MHz) & &Shift from  $^{173}$Yb\,[$\frac{5}{2}\rightarrow\frac{7}{2}$] (MHz) \\
   \vspace{-0.3cm}  \\
  \hline
     \vspace{-0.3cm}  \\
 Transition       & This work 	& Ref.~\cite{Pan2009} & Ref.~\cite{Wij1994} & This work \\
 \hline
    \vspace{-0.3cm}  \\
$^{173}$Yb\,($\frac{5}{2}\rightarrow\frac{7}{2}$)  & -1431.363\,(37) & -1431.872\,(60) & -1432.6\,(12) &  0  \\		
$^{171}$Yb\,($\frac{1}{2}\rightarrow\frac{1}{2}$)  & -1176.725\,(33)   & -1177.231\,(60)  & -1177.3\,(11) & 254.637\,(10) \\			
$^{176}$Yb & 0  &  0 & 0 & 1431.363\,(37)\\						
$^{174}$Yb & 954.628\,(26) & 954.832\,(60) &954.6284\,(9) & 2386.011\,(39)  \\			
$^{172}$Yb	& 1955.444\,(24) &  1954.852\,(60)  &1954.8\,(16) & 3386.827\,(38)\\	
$^{170}$Yb	& 3241.229\,(58) & 3241.177\,(60) & 3241.5\,(28) & 4672.612\,(64) \\
$^{173}$Yb\,($\frac{5}{2}\rightarrow\frac{5}{2}$) &  3266.554\,(61) & 3266.243\,(60) & 3267.1\,(28) &  4697.920\,(35) \\
$^{168}$Yb	&- & 4609.960\,(80) & 4611.9\,(44) &-\\
$^{171}$Yb\,($\frac{1}{2}\rightarrow\frac{3}{2}$) & 4761.175\,(44) & 4759.44\,(80) & 4761.8\,(37) & 6192.416\,(56)  \\
$^{173}$Yb\,($\frac{5}{2}\rightarrow\frac{3}{2}$) & 4763.680\,(300) & 4762.11\,(120) & 4761.8\,(37) & 6194.920\,(300)\\
   \vspace{-0.3cm}  \\
   \hline  
 \end{tabular}
 \end{table*}
 
 In previous work the transfer of atoms to the 556\,nm MOT is sometimes carried out with a reduction in the magnetic field gradient, $\mathrm{d}B/\mathrm{d}z$.  In our case a reduction in $\mathrm{d}B/\mathrm{d}z$ reduces the transfer efficiency. The magnetic field gradient of 0.42\,T$\cdot$m$^{-1}$ for the green MOT, at first, appears surprisingly high.   We can estimate the maximum $B$-gradient by considering the adiabatic condition that $\mathrm{d}B/\mathrm{d}z<\hbar k \Gamma /2 m v \mu$, where $k$ is the wave number, $m$ the mass of the atom, $v$ the velocity and $\mu$ is the Zeeman shift coefficient~\cite{Phi1982,Kat1999}.   For  the \coolingT\ transition  in \Yb\, ($F'=3/2$), $\mu=2 \pi\times 0.7\times10^{10}$\,Hz$\cdot$T$^{-1}$.   With a first stage temperature of \si300\,\uK\ ~\cite{Kos2014} \textit{all} the atoms can in principle be transferred with a capture velocity of about 0.6\,m\,s$^{-1}$.   The corresponding maximum $B$-field gradient is 1.0\,T$\cdot$m$^{-1}$, safely above what we find in the experiment.

\section{Isotopic shifts and hyperfine separations}  \label{Isotope}

The saturated absorption scheme provides a convenient means of measuring isotopic frequency shifts between Yb isotopes and the hyperfine separations within the fermionic isotopes.    To our knowledge, sub-Doppler spectroscopy with quantitative measurements has not been carried out previously on the \coolingT\ Yb lines.  The frequency comb (referred to above) is used to measure absolute optical frequencies \wrt\ a hydrogen maser for all the lines.    For each transition  [apart from $^{173}$Yb\, $(F=\frac{5}{2}\rightarrow\frac{3}{2})$ and $^{168}$Yb]  the 556\,nm light is locked to the center of the saturated absorption dip and the frequency of the optical comb beat is recorded with a frequency counter, set to a gate time of 0.4\,s, over a mean duration of 200\,s.   The Allan deviation typically reduces to \si10\,kHz with the accumulated data.   Our reported values in Table~\ref{isoshifts}  are the mean from six separate measurement runs taken over the course of two months.  During a single run the absolute frequencies are measured within a two hour period, hence factors that could affect the absolute frequencies, such as slight beam direction changes,  are made common to all the lines.  
 We evaluate the frequency separations with respect to the $^{176}$Yb isotopic line to aid comparison with previous work.
 For convenience, in the right most column we list the frequency separations \wrt\ the lowest frequency line:  $^{173}$Yb\, $(F=\frac{5}{2}\rightarrow\frac{7}{2})$.  These measurements were performed with a nulled magnetic field avoiding separation of the Zeeman components and the optical pumping behaviour described above.   We note that for the  \Yb\ (F'=3/2) line there is a \si200\,kHz  shift between the center of the enhanced absorption line and the zero-$B$-field line, while for the  \Yb\ ($F'=1/2$) line this shift is below 40\,kHz. 


Our quoted uncertainties are the statistical uncertainties ($\sigma/\sqrt{n-1}$) produced over the $n=6$ separate measurement trials, rather than an evaluation of systematic shifts.  
This is done because  the largest uncertainty arises from identifying the line-center for the lock point, rather  than from systematic shifts.  Note, even in the case of line-center offsets these will be largely common between the different lines as the lock point on the discriminator is not changed during a single measurement run (though, prior to the frequency measurement, a scan of the line was in most cases carried out to confirm symmetry about the lock point). 
There will, however, be an associated shift due to the different line strengths, which we estimate to be below 30\,kHz when considering the strongest and weakest lines.  For the most abundant isotopes the gain of the PMT was reduced to help minimise this shift.    A statistical uncertainty over a number of measurements should incorporate errors associated with identifying the line-center.  
Shifts due to the Stark effect or frequency counting errors  are well below the uncertainties reported here (and would also be common to all the lines).   In support, a 556\,nm beam intensity increase by a factor of ten showed no measurable  change to the (absolute) frequencies. 

 Some of the frequency separations differ by more than two standard deviations in comparison to previous measurements; notably the separations to the $^{173}$Yb\, $(F'=7/2)$, $^{171}$Yb\, $(F'=1/2)$ and $^{172}$Yb lines. In the case of $^{171}$Yb\, $(F'=3/2)$  we have better agreement with Ref.~\cite{Wij1994}.  We do not have an explanation for these differences.  As noted earlier, the $^{171}$Yb\, $(F'=3/2)$ line  is neighboured with the $^{173}$Yb\, $(F'=3/2)$ line.  The latter is only marginally visible with a +2.5\,(3)\,MHz separation from the former. 

\section{Conclusions}


We have identified an inverted crossover resonance in a divalent atom by use of saturated absorption spectroscopy where the transitions are driven with linearly polarised  retro-reflected beams. 
   It occurs with the two hyperfine lines in neutral $^{171}$Yb when a small magnetic bias field is applied parallel to the light's $\textbf{\emph{E}}$-polarisation.  The forward and reverse  beams are overlapped and orthogonal to a thermal atomic beam.  
The quality factor of the crossover resonance is \si2\ee{8}.   
A small magnetic bias field creates a simple optical pumping scheme that produces enhanced absorption via Zeeman splitting of the $J=0$ ground state.   
Branching to a ground state doublet produces multiple pathways for the atoms through the four states, creating a scheme free of saturation. 
  The photon absorption only becomes limited by the interaction time. 

 The $F=1/2$ energy level structure  of the \Yb\ intercombination line is not common; two other examples are $^{199}$Hg and $^{111,113}$Cd, partly explaining why an inverted crossover resonance has rarely (if at all) been observed in divalent atoms. 
An inverted crossover resonance is also observed with $^{173}$Yb, where $I=5/2$, but with a much poorer contrast and broader line shape.  
This is most likely because the multi-level splitting encompasses a broader velocity range, thus preventing the formation of a narrow resonance. 
Crossover resonances are further observed for the bosonic isotopes when the light's $\textbf{\emph{E}}$-polarisation  is perpendicular to the magnetic field. In this case the resonance forms a normal saturation feature. 


A dispersive signal derived from the enhanced absorption resonance is used to frequency stabilise 556\,nm light and in so doing enable laser cooling of \Yb\ atoms in a dual-stage \MOT.  Atom-cloud temperatures of \si20\,\uK\ are produced with cycle-to-cycle variations  of 2\,\uK.   This is achieved without the use of an optical reference cavity for high-frequency noise suppression.  The saturated absorption scheme has also provided the highest resolution measurements to date of the isotopic shifts and hyperfine separations  of the  $(6s^{2})$~$^{1}S_{0}$ $-$  $(6s6p)$\,$^{3}P_{1}$  transition in ytterbium.  


\section*{Acknowledgement}
This work was supported by the \ARC\ (grant LE110100054, M.E. Tobar).  J.M. is supported through an ARC Future Fellowship (FT110100392).  We express thanks to E.~de~Clercq, N.~Nemitz, D. Akamatsu, R. Anderson, and A. Gordon for helpful discussions, and L. Salter for assisting with the frequency measurements.  We are grateful to the UWA technicians, G. Light and S. Osborne, for fabricating components of the experiment.   We thank  M.E. Tobar, S. Parker and E. Ivanov for the use of equipment and we appreciate having the loan of equipment from  D. Sampson and D. Lorenser of OBEL, UWA. We thank L. Nenadovi\'{c} for developing the control and data acquisition software.  


\begin{thebibliography}{10}
\bibitem{comment1}
As found through tests of our own.
\bibitem{comment2}
This was not used permanently due to a lack of analog output channels on the
  control hardware.
\end{thebibliography}

\end{document}